\def\c{\chi}\def\d{\delta}\def\e{\epsilon}

\def\m{\mu}\def\o{\omega}\def\q{\psi}
\def\y{\eta}

\def\de{\partial}
\def\id{\equiv}\def\mo{{-1}}\def\ha{{1\over 2}}
\def\qu{{1\over 4}}\def\di{{\rm d}}

\def\st{spacetime }
\def\fe{field equations }
                                   
\def\bg{background }\def\gs{ground state }

\def\rep{representation }
\def\ssy{spherically symmetric }
\def\cur{curvature }\def\tor{torsion }
\def\min{Minkowski }\def\ads{anti-de Sitter }
\def\poi{Poincar\'e }
\def\des{de Sitter }
 
\def\GR{general relativity }\def\GB{Gauss-Bonnet }
\def\EH{Einstein-Hilbert }

\def\section#1{\bigskip\noindent{\bf#1}\smallskip}
\def\nota{\footnote{$^\dagger$}}
\font\small = cmr8

\def\PL#1{Phys.\ Lett.\ {\bf#1}}\def\CMP#1{Commun.\ Math.\ Phys.\ {\bf#1}} 
\def\PRL#1{Phys.\ Rev.\ Lett.\ {\bf#1}} 
 
\def\NP#1{Nucl.\ Phys.\ {\bf#1}}\def\GRG#1{Gen.\ Relativ.\ Grav.\ {\bf#1}}

\def\PRep#1{Phys.\ Rep.\ {\bf#1}}

\def\ref#1{\medskip\everypar={\hangindent 2\parindent}#1}
\def\beginref{\begingroup
\bigskip
\centerline{\bf References}
\nobreak\noindent}
\def\endref{\par\endgroup}

\def\eps{\e_{abcd}}\def\uab{(1+a^2-b^2)}\def\eab{(a\y^0-b\y^1-\qu x_k\y^k)}
\def\eu{{\y^1}}\def\ez{{\y^0}}\def\ei{{\y^i}}\def\eq{{\y^4}}
\def\cR{{\cal R}}\def\cT{{\cal T}}\def\cF{{\cal F}}
\def\dx{\,\di\hat x}\def\dr{\,\di r}\def\dt{\,\di t}

\def\uno{{(1)}}

\magnification=1200
\baselineskip18pt

{\nopagenumbers
\line{April 1997\hfil INFNCA-TH9704}
\vskip80pt
\centerline{\bf Spherically symmetric solutions in four-dimensional}
\centerline{\bf  Poincar\'e gravity with non-trivial torsion}
\vskip40pt
\centerline{{\bf S. Mignemi}\nota{e-mail:MIGNEMI@CA.INFN.IT}}
\vskip10pt
\centerline {Dipartimento di Matematica, Universit\`a di Cagliari}
\centerline{viale Merello 92, 09123 Cagliari, Italy}

\vskip40pt
{\noindent 
We study a four-dimensional gauge theory of the \poi group with topological
action which generalizes some well-known two-dimensional gravity models.
We classify the \ssy solutions and discuss the perturbative propagation of
excitations around flat spacetime.}
\vskip60pt
P.A.C.S. Numbers: 04.60.-m 04.70.Bw 11.15.-q
\vfil\eject}

\section {1. Introduction}
Two-dimensional models of gravity based on the gauge theory of the \poi
group or one of its generalizations have been extensively studied in recent
years [1]. The main reason for this interest resides in the fact that
two-dimensional models are much easier to handle than four-dimensional ones,
in particular in what concerns the issue of quantization and can therefore be
used as toy models of four-dimensional general relativity.

However, lower-dimensional models differ in several respects from the
four-dimensional gravity that they should imitate. For example, no propagating
degree of freedom is present in the spectrum. This fact can
be ascribed to the topological nature of the gravitational action in two
dimensions. One may wonder if these peculiarities spoil the analogy with
higher-dimensional gravity.

For this reason, we find interesting to investigate some aspects of the most
direct generalization to four dimensions of the standard two-dimensional
models of gravity and discuss its differences from general relativity.
This generalization was introduced some time ago in
ref. [2], where it was shown that a gauge theory of the \poi group with action
of topological form analogous to that used in two dimensions,
can be defined also in four dimensions. This action contains a multiplet
of scalar fields in addition to the geometric variables and
is quadratic in the curvature and the torsion.

In a previous paper [3], we have studied the riemannnian sector of this model,
where the torsion was set to zero. In contrast with the two-dimensional
case, however, in four dimensions the vanishing of torsion is not a
consequence of the field equations and hence we can extend the previous
investigations to the case of non-trivial torsion. Of course, the elimination
of the constraint of zero torsion may enlarge the number of degrees of
freedom of the theory and change its spectrum.

In this paper, we study in the general case the \ssy solutions of the \fe
and the propagation of the excitations around flat space. The main results
of our investigations is that, in contrast with general relativity, where the
Birkhoff theorem states that there is a unique family of \ssy solutions,
a large class of solutions is avalaible, which in general may depend
on arbitrary functions of the radial coordinate.
Moreover, we show that, in spite of the richer structure of the theory with
respect to the riemannian limit of ref. [3], also in this case no propagation
of excitations takes place in \min spacetime.   

The paper is organized as follows: in section 2 we introduce the model and
write down the field equations. In section 3 we impose a \ssy ansatz and in
section 4 classify all the possible solutions with this symmetry. In section 5
we discuss the perturbative propagation around the flat solution, while in
the last section we make some final remarks.

\section{2. Gauge theories of gravity in 4 dimensions}
The four-dimensional \poi group is isomorphic to $ISO(1,3)$, with generators
$M^{AB}=\{M^{ab},M^{a4}\id P^a\}$, where $A,B=0,\dots,4$; $a,b=0,\dots,3$.
The generators satisfy the usual commutation relations
$$[M^{AB},M^{CD}]=h^{AC}M^{BD}-h^{AD}M^{BC}+h^{BD}M^{AC}-h^{BC}M^{AD}$$
with $h^{AB}=$ diag $(-1,1,1,1,0)$.

As in standard Yang-Mills theory, local invariance under the \poi group
can be enforced by introducing a gauge connection one-form $A^{AB}$
with field strength 2-form $F^{AB}=dA^{AB}+A^{AC}A^{CB}$.
A gauge-invariant action of topological form can then be constructed 
making use of the totally antisymmetric group invariant tensor
$\e_{ABCDE}$. As in all even-dimensional models, one must further
introduce a multiplet of scalar fields $\y^A$, in the fundamental \rep of
the gauge group $ISO(1,3)$.
The action can then be written as\nota{\small
In the following we shall adopt the notations of [4] and omit the wedge signs.}
[2]:
$$I=\int_{M_4}\e_{ABCDE}\ \y^AF^{BC}F^{DE}\eqno(1)$$

In order to make contact with gravitation, one can then make the
identifications
$A^{ab}=\o^{ab}$, $A^{a4}=e^a$, where $\o^a_{\ b}=\o^a_{\ b\m}\di x^\m$ and 
$e^a=e^a_\m\di x^\m$ are the spin connection and  vierbein 1-forms of
the four-dimensional manifold.
These identifications imply that $F^{ab}=R^{ab}$, $F^{a4}=T^a$, where 
$R^{ab}$ and $T^a$ are the curvature and the torsion 2-forms of the
4-dimensional manifold, which are defined respectively as:
$$\eqalign{R^a_{\ b}&=d\o^a_{\ b}+\o^a_{\ c}\o^c_{\ b}\cr
T^a&=de^a+\o^a_{\ b}e^b}\eqno(2)$$
and satisfy the Bianchi identities 
$$\eqalign{DT^a&\id dT^a+\o^a_{\ b}T^b=R^a_{\ b}e^b\cr
DR^a_{\ b}&\id(DR+\o R-R\o)^a_{\ b}=0}\eqno(3)$$
where $D$ denotes the covariant derivative.

In terms of the geometrical quantities, the action (1) takes the form
$$I=\int_{M_4}L=\int_{M_4}\eps\left[\y^4R^{ab}R^{cd}+4\y^aT^bR^{cd}\right]
\eqno(4)$$

For future reference, we remark that, alternatively, the curvature and the
torsion can be written in terms of tensors in
an orthonormal basis as
$$R^{ab}=\cR^{ab}_{\ \ cd}e^ce^d\qquad\qquad T^{a}=\cT^{a}_{\ bc}e^be^c$$
In this formalism, the action reads
$$\eqalign{I=4\int e\ d^4x[&\y^4(\cR_{abcd}\cR^{cdab}
-4\cR_{ab}\cR^{ba}+\cR^2)\cr
&+4\y^a(\cT^b_{\ cd}\cR^{cd}_{\ \ ab}-2\cT^b_{\ ac}\cR^c_{\ b}+
2\cT_c\cR^c_{\ a}-\cT_a\cR)]\cr}\eqno(5)$$
where $e=\det e^a_{\ \m}$ and $\cR^a_{\ b}=\cR^{ac}_{\ \ bc}$,
$\cR=\cR^a_{\ a}$, $\cT_a=\cT^b_{\ ab}$.
(Notice that the ordering of indices is essential).

In order to evaluate the field equations, we vary the action with respect
to the independent fields $e$, $\o$, $\y^a$ and $\y^4$.
From the definition of the curvature, it follows that a variation $\d\o$ of
the connection induces a variation of the curvature given by
$$\d R^{ab}=D\d\o^{ab}=(d\d\o+\o\d\o+\d\o\o)^{ab}\eqno(6)$$

Using (6), the variation of the lagrangian $L$ can be written as
$$\eqalign{\d L=\eps&[R^{ab}R^{cd}\d\y^4+T^bR^{cd}\d\y^a+
2(\y^4R^{ab}+2\y^aT^b)D\d\o^{cd}\cr &-4\y^ae^hR^{cd}\d\o^b_{\ h}
+4\y^aR^{cd}D\d e^b]\cr}\eqno(7)$$
One can rearrange the $\o$-variation by noticing that, by definition of
covariant derivative,
$$\eqalign{D(\eps\y^4R^{ab}\d\o^{cd})&\id d(\eps\y^4R^{ab}\d\o^{cd})=
\eps(d\y^4R^{ab}\d\o^{cd}+\y^4R^{ab}D\d\o^{cd})\cr
D(\eps\y^aT^b\d\o^{cd})
&\id d(\eps\y^aT^b\d\o^{cd})+\y^a\o_a^{\ h}\e_{hbcd}T^b\d\o^{cd}\cr
&=\eps(d\y^aT^b\d\o^{cd}+\y^aR^b_{\ h}e^h\d\o^{cd}+\y^aT^bD\d\o^{cd})\cr}
\eqno(8)$$
where we have made use of the Bianchi identities (3) and of the tensorial
properties of $\eps$. Similarly, for the $e$-variation,
$$\eqalign{D(\eps\y^aR^{cd}\d e^b)
&\id d(\eps\y^aR^{cd}\d e^b)+\y^a\o_a^{\ h}\e_{hbcd}R^{cd}\d e^b\cr
&=\eps(d\y^aR^{cd}\d e^b+\y^aR^{cd}D\d e^b)\cr}\eqno(9)$$

Substituting in (7) the values of $D\d\o$ and $D\d e$ obtained from (8) and (9)
and discarding the total derivatives, since the variation of the fields are
independent it follows that
$$\eqalign{&\eps R^{ab}R^{cd}=0\cr &\e_{mbcd}T^bR^{cd}=0\cr
&\e_{mbcd}D\y^bR^{cd}=0\cr &\e_{mncd}(d\y^4R^{cd}+2D\y^cT^d+2\y^cR^d_{\ h}e^h)
-\e_{mbcd}\y^bR^{cd}e_n+\e_{nbcd}\y^bR^{cd}e_m=0\cr}\eqno(10)$$
where we have defined $D\y^a\id d\y^a+\o^a_{\ h}\y^h$.

\section{3. The field equations}
In the following, we shall be interested in static, \ssy solutions of the
field equations. The most general \ssy ansatz, which is also invariant under
reflections can be written as ($i,j,k=2,3$)\nota{\small
A more general ansatz can be obtained if one does not require reflection
invariance. In this case $\o$ can depend on eight independent functions
instead of four [5].}:
$$\eqalign{e^0&=f(r)\dt\cr e^1&=g(r)\dr\cr e^i&=r\dx^i\cr
\o^{01}&=c(r)\dt+d(r)\dr\cr \o^{0i}&=a(r)\dx^i\cr \o^{1i}&=b(r)\dx^i\cr
\o^{ij}&=\ha(x^i\dx^j-x^j\dx^i)\cr}\eqno(11)$$
where $r$ is the radial coordinate and $\dx^i=(1+{x^kx_k\over4})^\mo\di x^i$.
Moreover, we shall assume that $\y^a$ and $\y^4$ depend only on $r$.

In terms of these variables, the curvature and torsion 2-forms are:
$$\eqalign{R^{01}&=c'\dr\dt\cr R^{1i}&=(b'+ad)\dr\dx^i+ac\dt\dx^i\cr
R^{0i}&=(a'+bd)\dr\dx^i+bc\dt\dx^i\cr R^{ij}&=(1+a^2-b^2)\dx^i\dx^j\cr
T^0&=(f'-cg)\dr\dt\cr T^1&=df\dr\dt\cr
T^i&=(1+bg)\dr\dx^i-af\dt\dx^i\cr}\eqno(12)$$
where a prime denotes derivative with respect to $r$.

In order to obtain the \ssy solutions, one must now
substitute these expressions into the field equations (10). The independent
equations so obtained are listed below:
$$\eqalignno{
&\uab c'+2(aa'-bb')c=0&(13.a)\cr
&\uab df+2ac(1+bg)+2a(b'+ad)f=0&(13.b)\cr
&\uab(f'-cg)+2bc(1+bg)+2a(a'+bd)f=0&(13.c)\cr
&\uab(\eu'+d\ez)+2(b'+ad)\eab=0&(13.d)\cr
&\uab(\ez'+d\eu)+2(a'+bd)\eab=0&(13.e)\cr
&[\uab\ez+2a\eab]c=0&(13.f)\cr
&[\uab\eu+2b\eab]c=0&(13.g)\cr
&ac\ei'=bc\ei'=0&(13.h,i)\cr
&(a^2-b^2)c\ei=0&(13.j)\cr
&c'\ei=0&(13.k)\cr
&[a(b'+ad)-b(a'+bd)]\ei=0&(13.l)\cr
&(ac\ez)'=(bc\eu)'&(13.m)\cr
&(f'-cg)\ei=df\ei=0&(13.n,o)\cr
&(a^2f+rbc)\ei=a(bf+rc)\ei=0&(13.p,q)\cr
&[r(a'+bd)-a(1+bg)]\ei=[r(b'+ad)-b(1+bg)]\ei=0&(13.r,s)\cr
&bc\eq'-af\ez'-(bdf-acg+a'f+af')\ez-(3bcg+adf+c-bf')\eu=0&(13.t)\cr
&ac\eq'-af\eu'-(acg-bdf)\eu-(3adf+bcg+c+b'f)\ez=0&(13.u)\cr
&af\ei'-2(acg+bdf+a'f)\ei=0&(13.v)\cr
&[\uab\ez+2a\eab]f=0&(13.w)\cr
&\uab(\eq'-g\eu)-2(1+bg)\eab=0&(13.x)\cr
}$$

\section{4. Spherically symmetric solutions}
In spite of the huge number of equations to be solved,
it turns out that in general
they are not sufficient to determine uniquely all the eleven variables
$a, b, c, d, f, g, \y^a, \y^4$,
but almost all solutions depend on arbitrary functions. This is similar to
what occurs in the riemannian limit [3], where in general the solutions
depend on one arbitrary function.

In the following, we shall be interested only in the solutions with
$f, g\ne0$, since
they can be given a geometrical interpretation in terms of spacetime, but
it must be stressed that one
can also find many solutions with vanishing metric functions, which may be
interesting from the point of view of topological field theory.

One can make some general considerations about the solutions of the system
(13). If $\y^i$ vanishes, many of the \fe are
authomatically satisfied. The case $\ei\ne0$ is therefore quite special.
Let us examine it in more detail:
in the hypothesis $f\ne0$, eqs. ($13.n,o$) imply that $f'=cg$ and $d=0$.
Moreover, if $\ei$ does not depend on $x^k$, the terms containing $x_k\y^k$
in the \fe must vanish. This is possible iff $a=0$, $b =$ const $\ne0$,
$g=-b^\mo$ and $c=0$, which in turn implies $f$ = const. The remaining
equations admit two solutions: either $b^2$=1 (case A.1), or $b^2\ne1$ with
$\ez=\eu'=0$, $\eq'=\eu/b=$ const (case A.2).
The first case is the most interesting since it corresponds to flat \st
with vanishing torsion.

Let us consider now the case $\ei=0$. First, we assume that $1+a^2-b^2$ does
not vanish. In this case, eq. ($13.a$) can be integrated to yield
$$c={A\over1+a^2-b^2}\eqno(14)$$
with $A$ an arbitrary constant. (The possibility of integrating this
equation is a consequence of the fact that the \GB term in the action
is a total derivative in four dimensions). 
Moreover, if $c\ne0$, eqs. ($13.f,g$) can be combined as
$$(1+a^2-b^2)(b\ez-a\eu)=0\eqno(15.a)$$
$$(1+3a^2-3b^2)\ez=(1+3a^2-3b^2)\eu=0\eqno(15.b)$$
The equations (15) admit two different solutions, which we shall denote
B.1 and B.2: in the first case, $\ez=\eu=0$, while in the second case,
$1+3a^2-3b^2=0$ and $b\ez=a\eu$.
In case B.1, all equations except ($13.b,c$) are authomatically satisfied
if $\eq$= const; using the remaining equations one can hence determine two
functions, say $b$ and $d$, in terms of the others
and is left with three arbitrary functions $a$, $f$ and $g$. 

In case B.2, the solutions are harder to find. 
Combining ($13.d,e$) with ($13.f,g$) one can show that $\ez=Ba$, $\eu=Bb$,
with $B$ an integration constant. Substituting this result in ($13.x$)
one gets $\eq'=-B$.
With these values for the scalar fields, making use of the condition
$1+3a^2-3b^2=0$, eqs. ($13.b,c,t,u$) reduce to three independent equations,
which can be written as
$$af'-acg-bdf=0\qquad\qquad bf'-bcg-adf=0\eqno(16.a,b)$$
$$bf'+b'f+c=0\eqno(16.c)$$
Taking into account the relation between $a$ and $b$, the first two equations are
solved by $df=f'-cg=0$, while integrating the third after noticing that, due to
(14), $c$ = const = $3A/2$, one gets
$$f=-{1\over b}\left({3A\over2}r+C\right)\eqno(17)$$ 
with $C$ an integration constant.

We pass now to consider the case in which $1+a^2-b^2=0$, which we call B.3.
In this case eq. ($13.a$) is satisfied for arbitrary $c$. Moreover,
if $c\ne0$, eqs. ($13.f,g$) imply $a\ez=b\eu$. Making use of these conditions,
one is left with only two independent equations, which can be written as
$$d=-{c(1+bg)\over af}-{a'\over b}\qquad\qquad bc\eq'=af\ez'+
\left({c(b+g)\over a}-a^2a'f\right)\ez\eqno(18)$$
Therefore it results that in this case the five functions
$a,c,f,g,\ez$ can be chosen arbitrarily.

There are some further special solutions to the \fe which correspond
to the vanishing of some of the functions $c, d, \dots$, but we shall not
discuss them here, since they do not exhibit any particularly relevant
feature.

In table 1 we have listed the solutions discussed above.
The entry "any" means that the corresponding function can be chosen freely,
while we have
denoted by $\cF(\dots)$ the functional dependence on other variables implied
by the field equation (of course there is some arbitrarity in choosing which
variables are independent).

A common characteristic of all the solutions except B.1 is that some of
the components
of the \cur or the \tor vanish. In table 2 we have listed the vanishing
components for the given solutions.
Some special cases have physical relevance. For example, if one imposes
vanishing torsion, one can recover the solutions of ref. [3]. In fact,
vanishing torsion implies $a=d=0$, $c=f'/g$, $b=-1/g$. Substituting in the \fe
one obtains four possible classes of solutions (here we disregard the scalar
fields):

1) arbitrary $f$, $g^2=1$;

2) arbitrary $g$, $f=$ const;

3) $f=Ar+B$, $g^2=3$;

4) arbitrary $g$, $f=A(g^2-1)/g$;

{\noindent These are exactly the solutions found in ref. [3].}

Also interesting is the existence of solutions with vanishing curvature,
but non-trivial torsion. These are obtained whenever
$b^2=1+a^2, c=0, d=-a'/b$ and are therefore a special case of the solutions
B.3. No further restrictions are imposed on $f$ and $g$ by the field equations.

\section{5. Perturbative degrees of freedom}
The study of the propagation of excitations around a given \gs is useful
in order to investigate the physical content of the theory, and in particular
the number of degrees of freedom. Due to the topological nature of the model
we are considering, we expect no propagating degrees of freedom to appear
in the spectrum.

This question can be investigated by evaluating the part of the action
quadratic in the perturbations of the fields around the ground state.
For our purposes, the most suitable ground state is of course
Minkowski \st, which we have shown to be a solution of the \fe corresponding
to $f=g=1$, $a=c=d=0$, $b=-1$ and vanishing \bg values of the curvature and
torsion. From the
general form of our solutions, we are also induced to choose $\bar\y^a=0$,
$\bar\y^4$ = const as \bg values for the scalars.

The calculations are most easily performed in an orthonormal frame in
cartesian coordinates. Expanding
$\o^{ab}_{\ \ \m}=\bar\o^{ab}_{\ \ \m}+\c^{ab}_{\ \ \m}$,
$e^a_{\ \m}=\bar e^a_{\ \m}+h^a_{\ \m}$, where $\bar\o^{ab}_{\ \ \m}=0$,
$\bar e^a_{\ \m}=\d^a_\m$ are the background values of the connection and
vierbein corresponding to \min spacetime, and
$\c$ and $h$ are small perturbations, one has at first order
$$\eqalign{\cR^\uno_{abcd}&=\de_d\c_{abc}-\de_c\c_{abd}\cr
\cT^\uno_{abc}&=\de_a h_{bc}-\de_c h_{ba}\cr}\eqno(19)$$
with $\c_{abc}\id\d^\m_c\c_{ab\m}$ and $h_{ab}\id\d^\m_b h^a_{\ \m}$.
We also expand the scalar multiplet as $\y^A=\bar\y^A+\q^A$.

Substituting the expansion into the action (5), one sees that the linear part
of both $e(\cR_{abcd}\cR^{cdab}-4\cR_{ab}\cR^{ba}+\cR^2)$ and
$e(\cT^b_{\ cd}\cR^{cd}_{\ \ ab}-2\cT^b_{\ ac}\cR^c_{\ b}+
2\cT_c\cR^c_{\ a}-\cT_a\cR)$ vanishes,
since these terms are quadratic in the torsion and the curvature, whose \bg
values are null. Moreover, using (19), one can check explicitly that the
quadratic part of both terms is a total derivative.
From these results is easy to see that since 
$\bar\y^4$ and $\bar\y^a$ are constant, the full
quadratic action is a total derivative and hence no propagation arises around
flat space.

The absence of propagating degrees of freedom confirms the topological nature
of the theory. 
Of course, choosing a different \gs, in principle one could find different
properties for the propagation. A determination of the true degrees of freedom
can be better performed in a hamiltonian framework.

\section{6. Final remarks}
It would be straightforward to extend our investigations to the case of \des
or \ads groups. This has been done in [3] for the riemannian limit.
However, we do not expect any qualitatively new feature to emerge from this
generalization, even if in these cases the action contains also terms of the
\EH form.

A more promising development would be the investigation of the theory in the
hamiltonian formalism. As in two dimensions, this should give interesting
informations on the theory and its quantization and permit a rigorous
determination of its degrees of freedom.

We finally remark that different versions of four-dimensional gravity with
actions of topological type have been proposed [6]. These do not contain
scalar fields and hence are in some sense closer to \GR than the model
considered here.

\beginref
\ref [1] T. Fukuyama and K. Kamimura, \PL{B160}, 259 (1985);
K. Isler and C. Trugenberger, \PRL{63}, 834 (1989);
A.H.. Chamseddine and D. Wyler, \PL{B228}, 75 (1989);
D. Cangemi and R. Jackiw, \PRL{69}, 233 (1992);
\ref [2] A.H. Chamseddine, \NP{B346}, 213 (1990);
\ref [3] S. Mignemi,"Black hole solutions in four-dimensional topological
gravity", preprint INFNCA-TH9624;
\ref [4] B. Zumino, \PRep{137}, 109 (1986);
\ref [5] R. Rauch, J.C. Shaw and H.T. Nieh, \GRG{14}, 331 (1982);
\ref [6] G.T. Horowitz, \CMP{125}, 417 (1989);
Y.N. Obukhov and F. Hehl, Acta Phys. Pol. {\bf B27}, 2685 (1996).
\endref
\vfil\eject

\hoffset=-0.5in
\def\noah{\noalign{\hrule}}\def\qvad{$\;$}\def\ef{{\cal F}}
\def\eu{\y^1}\def\ez{\y^0}
\nopagenumbers
\halign{\strut#&\vrule\hfil\qvad#\hfil\qvad
&\vrule\hfil\qvad#\hfil\qvad
&\vrule\hfil\qvad#\hfil\qvad
&\vrule\hfil\qvad#\hfil\qvad
&\vrule\hfil\qvad#\hfil\qvad
&\vrule\hfil\qvad#\hfil\qvad
&\vrule\hfil\qvad#\hfil\qvad
&\vrule\hfil\qvad#\hfil\qvad
&\vrule\hfil\qvad#\hfil\qvad
&\vrule\hfil\qvad#\hfil\qvad
&\vrule\hfil\qvad#\hfil\qvad\vrule\cr
\noalign{\hrule}
&&$a$&$b$&$c$&$d$&$f$&$g$&$\y^0$&$\y^1$&$\y^i$&$\y^4$\cr\noah
&A.1&0&$\pm1$&0&0&const&$\mp1$&any&any&any&any\cr\noah
&A.2&0&const&0&0&const&$-b^\mo$&0&const&any&$Ar+B$\cr\noah
&B.1&any&$\ef(a,f,g)$&$A(1+a^2-b^2)^\mo$&$\ef(a,f,g)$&any&any&0&0&0&const\cr\noah
&B.2&any&$\pm\sqrt{a^2+1/3}$&const&0&$\ef(a,c)$&$\ef(a,c)$&$Ba$&$Bb$&0&$-Br+D$\cr\noah
&B.3&any&$\pm\sqrt{a^2+1}$&any&$\ef(a,c,f,g)$&any&any&any&$a\ez/b$&0&$\ef(a,c,f,g,\ez)$\cr\noah
\noalign{\smallskip}&\multispan 8 Table 1: Classification of the solutions
of the field equations.\hfil\cr
}
\vskip70pt
\halign{\strut#&\vrule\hfil\qvad#\hfil\qvad
&\vrule\hfil\qvad#\hfil\qvad
&\vrule\hfil\qvad#\hfil\qvad
&\vrule\hfil\qvad#\hfil\qvad
&\vrule\hfil\qvad#\hfil\qvad
&\vrule\hfil\qvad#\hfil\qvad
&\vrule\hfil\qvad#\hfil\qvad
&\vrule\hfil\qvad#\hfil\qvad\vrule\cr
\noalign{\hrule}
&&$R^{01}$&$R^{0i}$&$R^{1i}$&$R^{ij}$&$T^0$&$T^1$&$T^i$\cr\noah
&A.1&0&0&0&0&0&0&0\cr\noah
&A..2&0&0&0&&0&0&0\cr\noah
&B.1&&&&&&&\cr\noah
&B.2&0&&&&&0&0\cr\noah
&B.3&&&&0&&&\cr\noah
\noalign{\smallskip}&\multispan8 Table 2: The vanishing components \cr
\noalign{\smallskip}&\multispan8 of $R$ and $T$ for the different classes\cr
\noalign{\smallskip}&\multispan8 of solutions.\hfil\cr
}
\end